\documentclass{aadebug}

\corr{0309027}{3}

\usepackage{pifont}             % Postscript characters
\usepackage{xspace}
\usepackage{amsmath}

\newcommand{\FAIL}{\text{\ding{56}}\xspace}
\newcommand{\UNRESOLVED}{\lower0.1ex\hbox{\epsfig{file=question.ps, 
      height=1.7ex}}}

\newcommand{\GCC}   {{\small GCC}\xspace}
\newcommand{\GNU}   {{\small GNU}\xspace}

\newcommand{\mathid}[1]{\text{\rmfamily\textit{#1}}}

\def\|#1|{\mathid{#1}}

\newcommand{\codeid}[1]{\text{\upshape\texttt{#1}}}

\def\<#1>{\codeid{#1}}

\definecolor{grey}{gray}{.5}
\definecolor{black}{gray}{0}

\begin{document}

\runningheads{Andreas Zeller}{Causes and Effects in Computer Programs}

\title{Causes and Effects in~Computer~Programs}

\author{
Andreas~Zeller\addressnum{1}
}

\address{1}{
Lehrstuhl f\"ur Softwaretechnik,
Universit\"at des Saarlandes,
Postfach 15\,11\,50,
66041 Saarbr\"ucken,
Germany.  
E-mail: zeller@acm.org,
Web: http://www.st.cs.uni-sb.de/$\sim$zeller/
}

\begin{abstract}
  Debugging is commonly understood as finding and fixing the cause of
  a problem.  But what does ``cause'' mean?  How can we find causes?
  How can we prove that a cause is a cause---or even ``the'' cause?
  This paper defines common terms in debugging, highlights the
  principal techniques, their capabilities and limitations.
\end{abstract}


\begin{figure}[t]
\includegraphics[width=\textwidth]{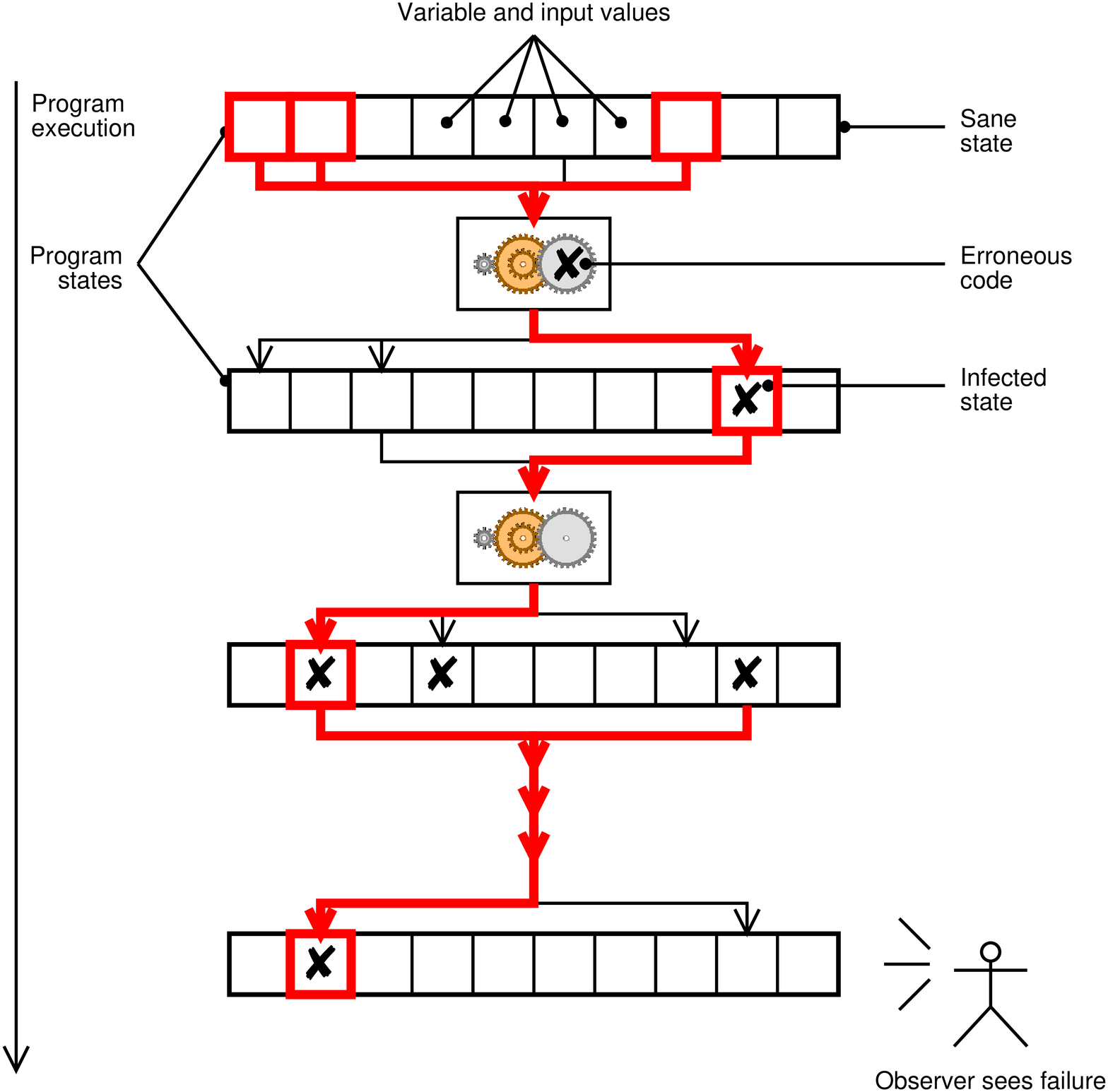}
\vspace{-0.5cm}
\caption{How failures come to be}
\label{fig:chain}
\end{figure}

\section{How Failures Come to Be}

Software bugs are a pain.  In the US alone, software bugs cause costs
of nearly 60 billion~US\$ anually, with more than half of the cost borne
by end users\cite{nist/2002/testing}.  How can we get rid of bugs?
Basically, there are two methods:
\begin{itemize}
\item We can \emph{prevent} bugs (or in other words: write correct
  software).  A plethora of tools and techniques is available to
  prevent bugs, including better languages, better processes, better
  analysis, formal verification.  But despite all the advances made in
  preventing bugs, it seems our programs just grow accordingly such
  that the number of bugs stays constant.
  
\item We can \emph{cure} bugs (or in other words: to fix the defect).
  This is the dark corner of computer science---very few people seem
  to care about debugging.  Obviously, to talk about prevention is
  nobler than examining the cure.  But the cure has been neglected in
  such a way that the process of identifying and correcting the root
  cause of a failure is still as labor-intensive, manual, and painful
  as it was 50~years ago.
\end{itemize}

\noindent
This paper focus on the \emph{cure} of bugs---that is, debugging.  But
what is it that makes debugging so difficult?  In principle, a failure
occurs in four steps, illustrated in Figure~\ref{fig:chain}:

\begin{description}
\item[1. A programmer creates a \emph{defect}\footnotemark in the
  code.] \footnotetext{The appendix of this paper contains a glossary
    for this and other important terms.}  A defect is a piece of the
  code that can cause an in\-fection---a part of the program state
  that can cause a failure.  In Figure~\ref{fig:chain}, the defect is
  shown by a \FAIL~character.\footnote{It is common to say that each
    defect is created by a \emph{mistake} made by the programmer.  However,
    there are several settings where it is difficult to blame an
    individual for a defect---and anyway, whom to blame is a
    political, not a technical question.}

\item[2. The input causes \emph{execution} of the defect.]  If the
  defect is not executed, it cannot trigger the failure in question.
  
\item[3. The defect causes an \emph{infection.}]  An infection is a
  part of the program state that can cause a failure; it is created by
  a defect being executed.  In Figure~\ref{fig:chain}, the infected
  states are shown by \FAIL~characters; you can see how the infection
  propagates along the execution.

  A defect in the code does not necessarily cause infections.  The
  defect code must be executed, and it must be executed under such
  conditions that the infection actually occurs.
  
\item[4. The infection causes a \emph{failure.}]  A failure is an
  externally observable error in the program behavior; it is created
  by an infection in the program state.

  An infection in the program state does not necessarily cause a
  failure.  It may be masked or corrected by some later program action
  before it could be observed.
\end{description}

\noindent
One must keep in mind that not every defect results in an infection,
and not every infection results in a failure.  Hence, having no
failures does not imply having no defects.  This is the curse of
testing, as pointed out by Dijkstra~\cite{dijkstra/72/ap}: it can only
show the presence of defects, but never their absence.  However,
\begin{itemize}
\item each failure can be traced back to an infection, 
\item each infection can be traced back to a defect that caused the
  infection,
\item each defect can be traced back to an input that caused the
  defect to be executed.
\end{itemize}

\noindent
Isolating this \emph{cause-effect chain} within the program run---from
input via the defect to the final state, as highlighted in
Figure~\ref{fig:chain}, is the main part of debugging.  The second,
smaller, part is to \emph{fix} the defect such that the failure no
longer occurs---that is, one must \emph{break} the cause-effect chain.

Why is finding a cause-effect chain so difficult?  First, there is a
lack of methodology in debugging.  In natural science, it is mandatory
to make the \emph{scientific method explicit}---that is, to write down
the hypotheses, the expectations, the experiments, and the conclusions
from these experiments.  If you don't write down what you're doing,
you must keep all these things in your head---which is why you cannot
leave your desk until the failure cause is found and fixed.

\begin{figure}[t]
\begin{center}
\includegraphics[width=\textwidth]{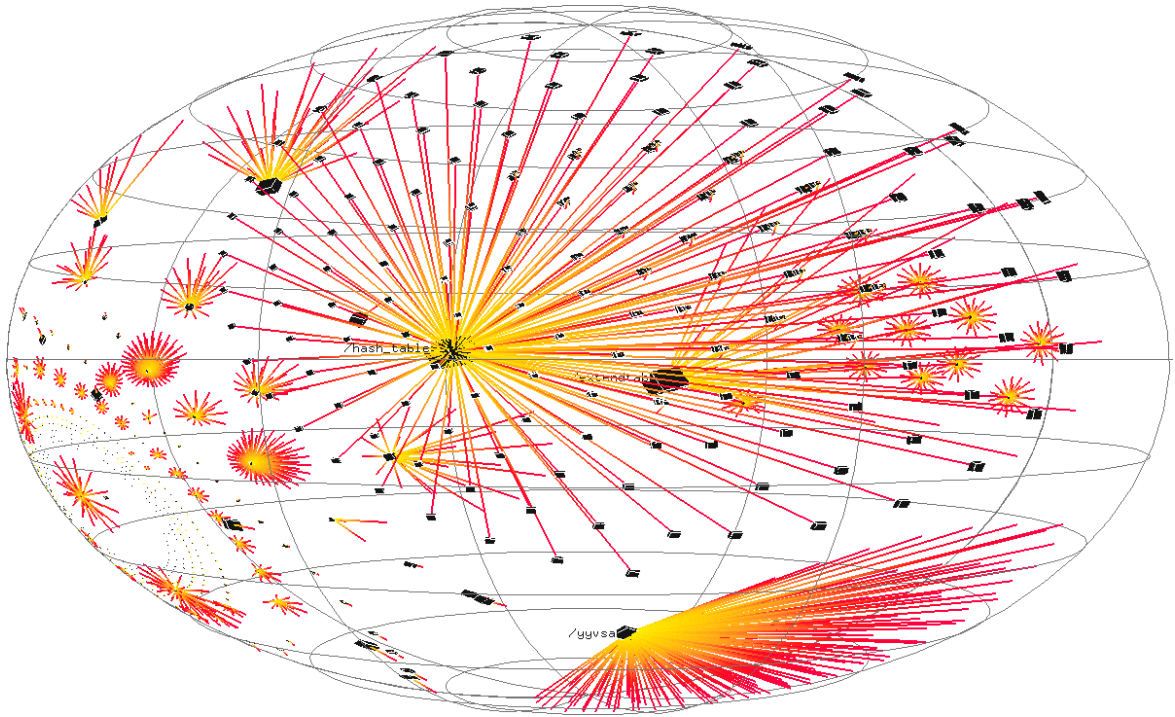}
\end{center}
\vspace{-0.5cm}
\caption{A program state of the \GNU C compiler}
\label{fig:gcc-fail}
\end{figure}

But even if you are conscious about bookkeeping, the \emph{search
  space} simply is enormous.  Program states are large, with tens of
thousands of variables.  Figure~\ref{fig:gcc-fail} shows the program
state of the \GNU C compiler with 42,991~variables (vertices) and
44,290~references (edges).  And this is only one single state at one
moment in time---an actual run is composed of millions to billions of
such program states.  Now, if tell you that \emph{one} single
reference in the shown program state causes the compiler to crash, how
do you find it?  You clearly need to be an expert in the program
you're analyzing in order to make good guesses about problem causes.

\section{About Causality}

To understand debugging, we must understand causes and effects.  If we
say that a defect \emph{causes} a failure, what does this actually
mean?  The most common definition for causality is as follows:

\begin{quote}
  A \emph{cause} is an event preceding another event without which the
  event in question would not have occurred.  The event in question is
  called the \emph{effect.}
\end{quote}
Hence, a defect causes the failure if the failure would not have
occurred without the defect.

In natural and social sciences, causality is often hard to establish.
Just think about common disputes such as ``Did usage of the butterfly
ballot in West Palm Beach cause George W. Bush to be President of the
United States?'', ``Did drugs cause the death of Elvis Presley?'', or
``Does human production of carbon dioxide cause global warming?''.

To determine whether these are actually causes, formally, we would
have to repeat history \emph{without the cause in question}---in an
alternate world that is as close as possible to ours.  If in this
alternate world, Albert Gore were president, Elvis were alive, global
warming were less, we knew that butterfly ballots, drugs, or carbon
dioxide had been actual causes for the given effects.

Unfortunately, we cannot repeat history.  We can only speculate, and
this is why anyone can always come up with a new theory about the true
cause, and this is why some empirical researchers have suggested
dropping the concept of causality altogether.

In our domain of computer science, though, things are different.  We
can easily repeat program runs over and over, change the circumstances
of the execution as desired, and observe the effects.  Given the right
means, the program execution is under total control and totally
deterministic.

Scientists frequently use computers to determine causes and effects in
\emph{models} of the real world.  However, such causes and effects run
the danger to be inappropriate in the concrete, because the model may
have abstracted away important aspects.  If we are determining causes
and effects in the program itself, though, there is no such risk---we
are already dealing with the real thing.  Hence, \emph{debugging is
  the only scientific discipline which can claim dealing with actual
  causality.}

However, the above definition of cause is not without problems.
Consider some failing program.  By the definition above, its entire
code is a cause for the failure (because without the entire code,
there would be no program to execute).  Likewise, the existence of
electricity is a failure cause, because without electricity\dots well,
you see how the argument goes.

\begin{figure}
\includegraphics*[width=\textwidth]{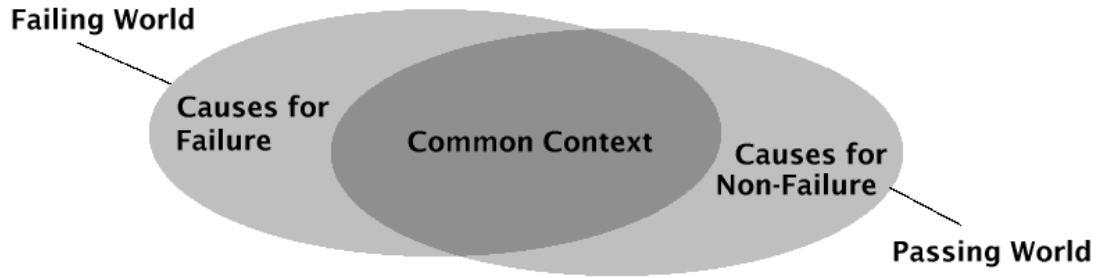}
\caption{Causes as differences between alternate worlds}
\label{fig:worlds}
\end{figure}

To discriminate between these alternatives, the concept of the
\emph{closest possible world} comes in handy.  A world is said to be
``closer'' to the actual world than another if it resembles the actual
world more than the other does.  The idea is that ``the'' cause should
be a \emph{minimal difference} between the actual world where the
effect occurs and the alternate world where it would not
(Figure~\ref{fig:worlds}).  In other words, the alternate world should
be \emph{as close as possible:}

\begin{quote}
  An \emph{actual cause} is a difference between the actual world
  where the effect occurs and the \emph{closest possible world} where
  it would not.
\end{quote}

\noindent
The concept of the closest possible world is applicable to all
causes---including the causes required for debugging.  So, if we want
to find the actual cause for a program failure, we can quickly
eliminate ``causes'' like the existence of electricity---a world
without electricity would be so much different from ours that it would
hardly qualify as ``closest''.  Instead, we have to search for the
\emph{closest possible world in which the failure does not occur:}

\begin{itemize}
\item The actual failure cause in a \emph{program input} is a minimal
  difference between the actual input (where the failure occurs) and
  the closest possible input where the failure does not occur.
\item The actual failure cause in a \emph{program state} is a minimal
  difference between the actual program state and the closest possible
  state where the failure does not occur.
\item The actual failure cause in a \emph{program code} is a minimal
  difference between the actual code and the closest possible code
  where the failure does not occur.
\end{itemize}

\noindent
The principle of picking the closer alternate world is also known as
\emph{Ockham's Razor}---whenever you have competing theories for how
some effect comes to be, pick the simplest.  As a side effect, this
quickly eliminates unlikely causes like ``the compiler doesn't work'',
``the CPU must be wrong'', ``Aliens tampered my computer''---or
whatever young programmers come up with to decline
responsibility.\footnote{Well, not quite.  In my 20-year career as a
  programmer, I actually found one defect in the 6502~CPU, two defects
  in the GNU compiler, and one defect in the C++ language definition.
  This sums up to one ``unlikely'' defect per five years---a reason to
  eliminate all other potential causes first.}

The definition of actual causes carries important side-effects for
debugging.  Most important is, you cannot tell that some property is a
failure cause until it has been altered such that the failure no
longer occurs.  In other words: \emph{The proof of the error is in the
  fix.}

\section{Reasoning About Programs}

So, how can one find failure causes?  Let me first line up some
reasoning techniques that do \emph{not} find failure
causes~\cite{zeller/2003/woda}:

\begin{description}
\item[Deduction.]  Deduction is reasoning from the abstract into the
  concrete.  In debugging, deduction techniques typically come in the
  form of \emph{static} analysis techniques that abstract from the
  (abstract) program code into the (concrete) program run.  They
  determine what can happen during a program run (or, more precisely,
  exclude what \emph{cannot} happen).  In debugging, the typical
  instance of static analysis is \emph{program
    slicing}~\cite{weiser/82/cacm, tip/95/jpl}.
  
  Deduction techniques are great in finding \emph{errors}---deviations
  from what's correct, right, or true.  Several syntactical and
  semantical errors can be caught by deduction.  So why can't
  deduction find failure causes?  To put things straight: deduction
  can find lots and lots of errors, and it deserves all our credits.
  
  However, to prove that an error (or a potential error) is an actual
  failure cause requires program execution, and deduction prohibits
  dealing with the concrete.  Furthermore, deduction works on an
  \emph{abstraction} of the real thing, and there is always a risk of
  some failure cause being abstracted away.
  
\item[Observation.]  Observation allows the programmer to inspect
  arbitrary aspects of an individual program run.  Since an actual run
  is required, the associated techniques are called \emph{dynamic.}
  Observation brings in actual \emph{facts} of a program execution;
  unless the observation process is flawed, these facts cannot be
  denied.

  Observation is the base of all interactive debugging tools, and has
  helped millions of programmers narrowing down defects.  But
  observation alone, again, cannot find failure causes.  You need to
  prove that \emph{altering} the suspected infection or defect
  actually causes the failure to disappear.

\item[Induction.]  Induction is reasoning from the particular to the
  general.  In program analysis, induction is used to \emph{summarize}
  multiple program runs---e.g. a test suite or random testing---to
  some abstraction that holds for all considered program runs.  The
  key concept added induction is \emph{anomalies}---the way a failing
  run differs from one or more working runs.
  
  Early uses if induction involved only two runs---a working and a
  failing one.  A \emph{relative
    debugger}~\cite{sosic/abramson/97/spe} compares two runs and
  reports any differences it can find.  A
  \emph{dice}~\cite{lyle/weiser/87/icam} or difference between two
  dynamic slices shows up the statements that were involved in
  producing the failure, but not in producing the correct
  result---these anomalies frequently include the defect.  Likewise,
  comparing the code coverage can highlight potentially erroneous
  code~\cite{jones/etal/2002/icse}.  \emph{Dynamic
    invariants}~\cite{ernst/etal/2001/tse} can be also be used to
  detect anomalous program behavior~\cite{hangal/lam/2002/icse}.

  Anomalies are not necessarily errors; they are not even failure
  causes unless verified as such.  But they can be excellent starting
  points when searching for errors and failure causes.
\end{description}

\begin{figure}[t]
\begin{center}
\includegraphics[width=0.8\textwidth]{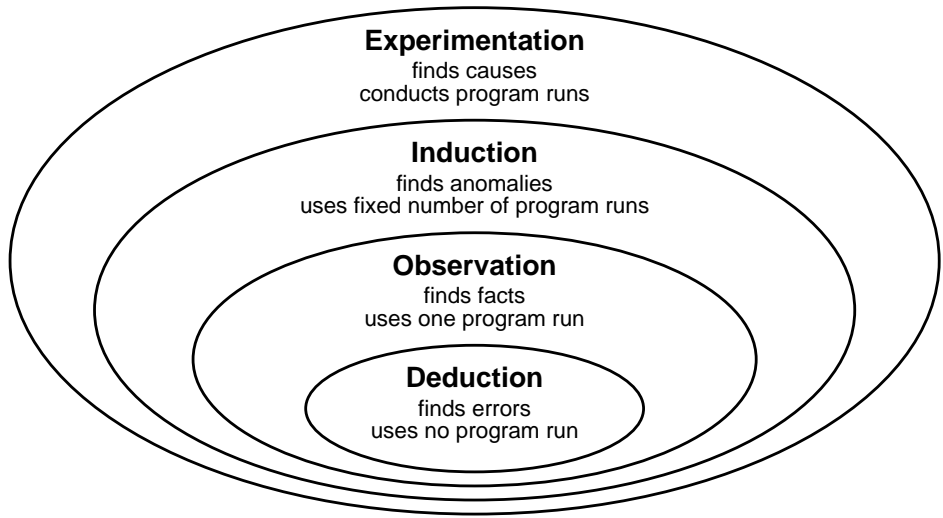}
\end{center}
\vspace{-0.25cm}
\caption{A hierarchy of reasoning techniques}
\label{fig:hierarchy}
\end{figure}

\noindent
So, what can we use to find failure causes?  The answer is the fourth
reasoning technique:

\begin{description}
\item[Experimentation.]  Experimentation means to \emph{conduct}
  experiments---typically, in order to verify or reject a specific
  hypothesis.  In our case, these experiments are program runs.  While
  induction is based on a \emph{fixed} set of program runs,
  experimentation adds and assesses new runs, specifically created for
  the hypothesis in question.
  
  Experimentation is the \emph{only} way to prove that an event is an
  actual cause---because you need two experiments for this.  While
  deduction, observation, and induction are all helpful in finding
  potential causes, only experimentation can separate the wheat from
  the chaff---and narrow down the actual cause of a failure.
  Experimentation is the essence of the \emph{scientific method}, the
  general process of constructing a theory for some aspect of the
  universe simply by hypotheses and experiments.  Experiments must be
  \emph{repeatable,} which in case of programs means that their
  execution must be made \emph{deterministic}---for instance, by means
  of a capture/replay tool~\cite{choi/srinivasan/98/spdt}.  Without
  determinism, there is no way to prove causality.

  Today, experimentation is still mostly conducted by humans, using
  all the reasoning techniques described above.  However, under
  certain conditions, experimentation can be \emph{automated,} thus
  effectively automating the search for failure causes.  This is the
  case when \emph{two} program runs exist---a working and a failing
  one---and the initial difference between these two runs can be
  narrowed down to isolate failure causes.

  The so-called \emph{delta debugging} approach has been shown to
  isolate failure-inducing input~\cite{zeller/hildebrandt/2002/tse},
  failure-inducing code changes~\cite{zeller/99/esec}, and
  failure-inducing program states~\cite{zeller/2002/fse}.  The process
  is based on an automated test that determines whether the failure
  persists or not, and an algorithm that systematically narrows down
  the difference between the two runs.  Applied to states, this
  effectively produces \emph{cause-effect chains} from the input to
  the failure, as the one shown in Figure~\ref{fig:chain}.
\end{description}

\noindent
To sum up, deduction finds \emph{errors,} observation finds
\emph{facts,} induction finds \emph{anomalies,} and experimentation
finds \emph{causes}---and most of these techniques can be automated.
The four reasoning techniques do not exist in isolation, though, nor
are they used at all as isolated techniques.  Each of the ``later''
techniques above can make use of ``earlier'' techniques.  For
instance, observation can make use of deduction when deciding what to
observe; Induction requires observation, and experimentation relies on
induction.

On the other hand, deduction \emph{cannot} use observation because
observation requires program execution, which deduction does not
allow; and observation cannot make use of induction, because
observation is based on one single run.  Consequently, these four
techniques form a \emph{hierarchy} (Figure~\ref{fig:hierarchy}), where
each ``outer'' technique can use elements of ``inner'' techniques, and
where each technique is defined by the number of program runs it uses
to make its findings.

\section{Finding ``the'' Error}

So, if experimentation can find failure causes automatically, and if
deduction can find errors, why don't we just fix all the bugs?  The
reason is subtle: \emph{Just because something is a cause, it need not
  be an error.}  If your computer catches a virus from an e-mail, the
virus code is the cause of the infection.  However, the virus code is
not an error---it is perfectly legitimate for e-mail to contain
arbitrary content.  The bad reference in Figure~\ref{fig:gcc-fail}, on
the other hand, is both a failure cause and an error, as it violates a
\GCC invariant.

What's an error, anyway?  Here's the definition:
\begin{quote}
  An \emph{error} is a deviation from what's correct, right, or true.
\end{quote}
Note that a cause might be an error or not; likewise, an error might
cause a failure (or the failure in question), or not.  The concepts of
errors and failure causes are \emph{totally unrelated}, except that,
of course, a failure is an error.  And while we can detect errors and
failure causes, automatically, it is hard to relate a failure to ``the''
error that caused the failure.

In practice, finding ``the'' error is not that difficult---provided
one has understood how the failure came to be, and this is where
observation, induction, and experimentation are most helpful.  In the
cause-effect chain (compare Figure~\ref{fig:chain}), we must find the
transition from sane (correct) states to infected (erroneous) states.
This transition is the defect, or the error in the code.  This simply
works by \emph{deciding} whether a state is sane or
infected---typically by inspection of the programmer.  

Isn't this something that could (and should) be fully automated, too?
Unfortunately, things are not so simple.  A \emph{cause} in a computer
program can be determined without doubt, using experiments---and
consequently, it can be determined automatically.  Whether some
property is an \emph{error,} though, can be formally decided if and
only if there is a \emph{formal specification} of what's correct,
right, or true.

A test case, for instance, decides whether a program is in error or
not.  A mere comment, or prose in a requirements document, can be
subject to interpretation, though.  In the absence of a formal
specification, there are no errors, only surprises; and we as
programmers must decide whether the surprise is desired or not.

The consequence is: As long as we can tell sane states from infected
states, we can narrow down the search quickly and even automatically,
but no further.  Once we have reached a cause-effect chain fragment
where such an automated assessment is impossible, the remaining search
is up to the user.  Note, though, that the programmer can be
\emph{guided} in her debugging process by querying whether a state is
sane or infected---the essence of \emph{algorithmic
  debugging}~\cite{shapiro/83/mit}.

Let us now assume we finally found ``the'' error, and we have proven
that it is a failure cause by altering it such that the failure goes
away.  Is this the way to go?  Not necessarily.  First, the change
applied to prove an error is an error is a fix for the failure in
question, but not necessarily a fix for other, related failures too.
One should opt for the \emph{most general fix} available---which may
involve rewriting not only ``the'' error, but other parts of the
program, too.  In some cases, the best fix is to rewrite the entire
program from scratch!  Since fixing is thus part of writing a program,
it is hard to see how automation could step in here.

There are other ``political'' issues about errors and fixing.  Suppose
you find that some function in a third-party library does not operate
as specified.  The third party says it's just your use of the library,
and the library won't be changed.  So, rather than changing the
erroneous code, you alter your original code to \emph{work around} the
problem.  Where's ``the'' error in that story?  Obviously, it depends
on whom you ask---and certainly you won't ask an automated tool here.

As researchers, we should be very precise about what our techniques
can do and what they cannot do.  This is especially important when
talking about anomalies, causes, and errors.  Yes, we do have
techniques that detect anomalies and errors in programs and runs.
Yes, for a given failure, we can isolate causes and cause-effect
chains automatically.  But a method that finds ``the'' error for a
particular failure---the ultimate target of automated debugging---must
prove that the error \emph{causes} the failure.  In other words, it
would have to provide not only the error, but the fix as well.  I see
no way of achieving this today.  I hope, though, that we'll be able to
detect so many anomalies and potential errors that we can leverage
this knowledge in the search for failure causes---and pinpoint the
error as good as we possibly can.

\section{Conclusion}

What's the future of automated debugging?  In this paper, I have not
attempted to cover the entire field of automated debugging, but rather
tried to group typical techniques of dynamic program analysis and
automated debugging---and to identify and relate their key concepts.
Today, it is obvious that the enormous wealth in computing power
brings clear benefits: We can afford computations, tests, and
experiments that would have seemed ridiculously large or dumb only a
few years ago.  In particular, the original
\begin{itemize}
\item deductive methods to determine errors and 
\item observational methods to determine facts
\end{itemize}
have now been adjoined by
\begin{itemize}
\item inductive methods to determine anomalies, and
\item experimental methods to determine causes. 
\end{itemize}
When debugging, the best programmers make use of all reasoning
capabilities---deduction, observation, induction, and experiments.
Our tools should do so, too.  The best tools will be those that
intertwine and dovetail errors, facts, anomalies, and causes.
Designing and building such tools will be the next challenge for
automated debugging.

\medskip
\noindent
\textbf{Acknowledgments.}  Holger Cleve and Stephan Neuhaus provided
substantial comments on earlier revisions of this paper.

\clearpage
\section*{Glossary}

\newcommand{\capitalize}[1]{\uppercase{#1}}
\newcommand{\caps}[1]{\expandafter\capitalize#1}
\newcommand{\glo}[1]{\item[\caps{#1}] \label{glo:#1}}
\newcommand{\see}[1]{$\rightarrow$\emph{#1}}
\newcommand{\Seealso}[1]{See also \emph{#1}}
\newcommand{\Aka}[1]{Also known as \emph{#1}}
\newcommand{\Compare}[1]{Compare \emph{#1}}

This glossary gives definitions for important terms as they are being
used in this paper.  Alternate definitions (2., 3., \dots) are being
found outside of this paper; references within the glossary always
refer to the first definition.

\begin{description}
\glo{accident} Synonym of \see{mishap}.

\glo{anomaly} A program behavior that deviates from expectations
  based on other runs or other programs.  \Aka{incident}.
  
\glo{bug} 1. Synonym of \see{defect}.  2. Synonym of \see{failure}.
  3. Synonym of \see{problem}.  4. Synonym of \see{infection}.

\glo{cause} An event preceding the \see{effect} without which the
  effect would not have occurred.

\glo{crash} The sudden and complete \see{failure} of a computer system
  or component.

\glo{debugging} Relating a \see{failure} or an \see{infection} to a
  \see{defect} and subsequent \see{repair} of the defect.

\glo{defect} An \see{error} in the program, esp. one that can cause an
  \see{infection} and thus a \see{failure}.  \Aka{bug, fault}.
  
\glo{delta} Difference between (or change to) code, states, or
  circumstances.

\glo{effect} An event following the \see{cause} that would not have
  occurred without the cause.

\glo{error} 1. An unwanted and unintended deviation from what is correct,
  right, or true.
  2. Synonym of \see{infection}.
  3. Synonym of \see{mistake}.

\glo{exception} An event that causes suspension of normal program operation.
  
\glo{failure} An externally visible \see{error} in the program
  behavior.  \Aka{malfunction}.  \Seealso{problem}.

\glo{fault} Synonym of \see{defect}.

\glo{feature} An intended property or behavior of a program.
  
\glo{fix} A \see{repair} where the \see{defect} is removed from the
  program.  \Seealso{debugging}.  \Compare{workaround}.

\glo{incident} Synonym of \see{anomaly}.

\glo{infection} An \see{error} in the program state, esp. one that
  can cause a \see{failure}.

\glo{issue} Synonym of \see{problem}.

\glo{malfunction} Synonym of \see{failure}.

\glo{mishap} An unplanned event or series of events resulting in
  death, injury, occupational illness, or damage to or loss of data
  and equipment or property, or damage to the environment.  \Aka{accident}.

\glo{mistake} A human act or decision resulting in an \see{error}.

\glo{problem} A questionable property or behavior of a program. 
  \Aka{issue}.  \Seealso{failure}.

\glo{repair} A \see{delta} such that the failure in question no longer
  occurs.  \Seealso{fix}.  \Seealso{workaround}.

\glo{testing} The execution of a program with the intention to find 
  some \see{problem}, esp. a \see{failure}.

\glo{test case} A documentation specifying inputs, predicted results,
  and a set of execution circumstances for a program.
  
\glo{workaround} A \see{repair} where the \see{defect} remains in the
  program.  \Compare{fix}.

\end{description}

\clearpage
\bibliography{debug,softech-e}

\end{document}